# Characterization of near-room-temperature superconductivity in yttrium superhydrides

E. F. Talantsev[1,2*]

[1]M.N. Mikheev Institute of Metal Physics, Ural Branch, Russian Academy of Sciences, 18, S. Kovalevskoy St., Ekaterinburg, 620108, Russia

[2]NANOTECH Centre, Ural Federal University, 19 Mira St., Ekaterinburg, 620002, Russia

[*]E-mail: evgeny.talantsev@imp.uran.ru

*Abstract*

Recently, Troyan *et al* (2019 arXiv:1908.01534) and Kong *et al* (2019 arXiv:1909.10482) extended near-room-temperature superconductors family by new yttrium superhydride polymorphs, $YH_n$ (n = 4,6,7,9), which exhibit superconducting transition temperatures in the range of $T_c$ = 210-243 K at pressure of $P$ = 160-255 GPa. In this paper, temperature dependent upper critical field data, $B_{c2}(T)$, for highly-compressed mixture of $YH_4+YH_6$ phases (reported by Kong *et al* 2019 arXiv:1909.10482) is analysed to deduce the ratio of $T_c$ to the Fermi temperature, $T_F$. Our analysis shows that in all considered scenarios the $YH_4+YH_6$ mixture has the ratio $0.01 \leq T_c/T_F \leq 0.04$. As the result, $YH_4+YH_6$ falls in the unconventional superconductors band in the Uemura plot. It is also found that the characteristic temperature of the order parameter amplitude fluctuations, $T_{fluc}$, in the $YH_4+YH_6$ mixture is only several percent above observed $T_c$, and thus the superconducting transition in yttrium superhydride polymorphs is fundamentally limited by thermodynamics fluctuations.

## I. Introduction

Recently, Troyan *et al* [1] and Kong *et al* [2] reported on the discovery of new superhydride polymorphs of yttrium, YH$_n$ (n = 4,6,7,9), which exhibit superconducting transition at $T_c$ = 210-243 K to be subjected to pressure of $P$ = 160-255 GPa. As far as pure hydrogen, pressurized to $P$ = 440 GPa, does not show any evidences for superconducting transition down to $T$ = 4.2 K [3], the near-room-temperature (NRT) hydrogen-rich superconductors family [4],[5],[6],[7] has been extended by new YH$_n$ polymorphs [1],[2]. Kong *et al* [2] attribute the highest observed $T_c$ = 237-243 K for *P6$_3$/mmc*-YH$_9$ phase, and this experimental findings are well supported by the first principles calculation studies [8],[9],[10],[11].

In this paper we analyse temperature dependent upper critical field data, $B_{c2}(T)$ (reported by Kong *et al* [2] for the mixture of YH$_4$+YH$_6$ phases) with the purpose to classify the superconductivity in this NRT superconductor and to deduce the characteristic temperature of thermodynamic fluctuations, $T_{fluc}$, in this highly-compressed superhydride of yttrium.

## II. Ground state coherence length

Kong *et al* [2] in their Figure 1(d) reported the temperature dependent magnetoresistance, $R(T,B)$, for Sample 5 pressurised at $P$ = 185 GPa, which contents the mixture of YH$_4$+YH$_6$ phases ($T_c$ ~ 214 K). By utilising the criterion of 75% of normal state resistance, $R(T)/R_{norm}$ = 0.75, we deduce the upper critical field, $B_{c2}(T)$, dataset and fit it to four models:

1. Gorter-Casimir model (GC model, Fig. 1(a)) [12]:

$$B_{c2}(T) = \frac{\phi_0}{2 \cdot \pi \cdot \xi^2(0)} \cdot \left(1 - \left(\frac{T}{T_c}\right)^2\right), \tag{1}$$

where $\phi_0 = 2.068 \cdot 10^{-15}$ Wb is magnetic flux quantum.

2. Gorkov model (Fig. 1(b)) [13]:

$$B_{c2}(T) = \frac{\phi_0}{2\cdot\pi\cdot\xi^2(0)} \cdot \left(\frac{1.77 - 0.43\cdot\left(\frac{T}{T_c}\right)^2 + 0.07\cdot\left(\frac{T}{T_c}\right)^4}{1.77}\right) \cdot \left[1 - \left(\frac{T}{T_c}\right)^2\right], \quad (2)$$

3. Werthamer-Helfand-Hohenberg model (WHH model, (Fig. 1(c)) [14]:

$$B_{c2}(0) = \frac{\phi_0}{2\cdot\pi\cdot\xi^2(0)} = -0.693 \cdot T_c \cdot \left(\frac{dB_{c2}(T)}{dT}\right)_{T\sim T_c}, \quad (3)$$

4. Baumgartner *et al* model (B-WHH model, Fig. 1(c)) [15]:

$$B_{c2}(T) = \frac{\phi_0}{2\cdot\pi\cdot\xi^2(0)} \cdot \left(\frac{\left(1-\frac{T}{T_c}\right) - 0.153\cdot\left(1-\frac{T}{T_c}\right)^2 - 0.152\cdot\left(1-\frac{T}{T_c}\right)^4}{0.693}\right). \quad (4)$$

Result of fits to these four models are shown in Fig. 1 and Table 1.

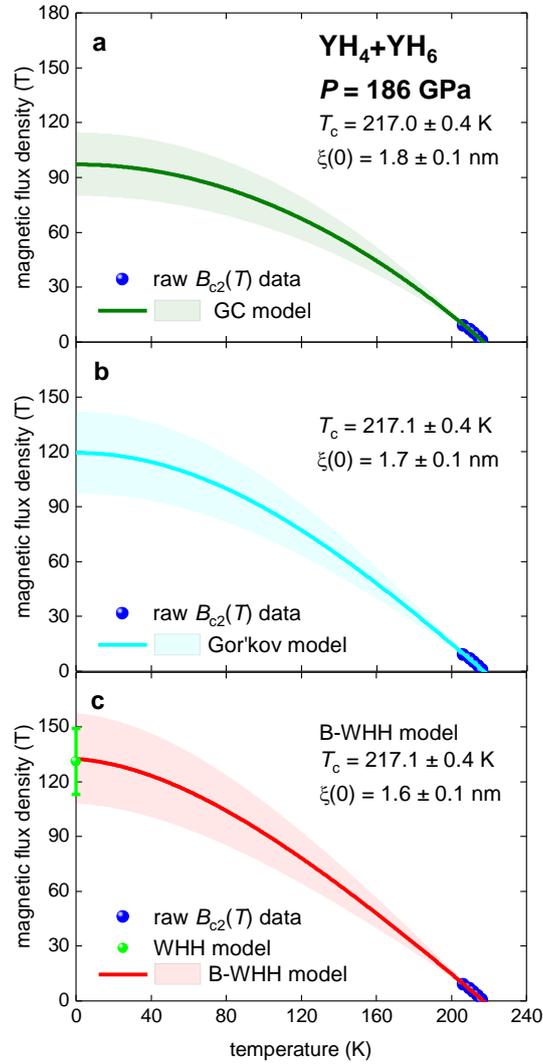

**Figure 1.** Superconducting upper critical field, $B_{c2}(T)$, data and fits to four models (Eqs. 1-4) for YH$_4$+YH$_6$ mixture (Sample 5 [2]) compressed at pressure of $P = 186$ GPa. (a) fit to GC model, the fit quality is $R = 0.982$. (b) fit to Gor'kov model, $R = 0.981$. (c) fit to WHH and B-WHH models, for latter $R = 0.981$. 95% confidence bars are shown.

**Table I.** Deduced and calculated parameters for the mixture of YH$_4$+YH$_6$ compressed at $P$ = 186 GPa (Sample 5 [2]). Deduced critical temperature for all models (Eqs. 1-4) is $T_c$ = 217 K. Assumed charge effective mass is $m^*_{eff} = 2.76 \cdot m_e$ [11]. Smallest and largest values for $\frac{T_c}{T_F}$, $\frac{T_c}{T_{fluc,phase}}$ and $\frac{T_c}{T_{fluc,amp}}$ are marked in bold.

| Model | Deduced $\xi(0)$ (nm) | Assumed $\frac{2 \cdot \Delta(0)}{k_B \cdot T_c}$ | $T_F$ (10$^3$ K) | $T_c/T_F$ | Assumed $\kappa$ | $T_{fluc,phase}$ (K) | $T_{fluc,amp}$ (K) | $T_c/T_{fluc,phase}$ | $T_c/T_{fluc,amp}$ |
|---|---|---|---|---|---|---|---|---|---|
| GC | 1.8 ± 0.1 | 3.53 | 7.2 ± 0.9 | 0.030 ± 0.003 | 60 | 3750 ± 190 | 1010 ± 55 | 0.058 ± 0.003 | 0.22 ± 0.01 |
|  |  | 5.47 | 17.4 ± 2.0 | **0.013 ± 0.02** | 120 | 940 ± 50 | 252 ± 13 | **0.23 ± 0.01** | **0.86 ± 0.05** |
| Gor'kov | 1.7 ± 0.1 | 3.53 | 6.5 ± 0.7 | 0.034 ± 0.004 | 60 | 3970 ± 220 | 1070 ± 60 | 0.055 ± 0.003 | 0.20 ± 0.02 |
|  |  | 5.47 | 15.5 ± 1.9 | 0.014 ± 0.001 | 120 | 994 ± 45 | 267 ± 15 | 0.22 ± 0.01 | 0.81 ± 0.05 |
| B-WHH | 1.6 ± 0.1 | 3.53 | 5.7 ± 0.9 | **0.038 ± 0.003** | 60 | 4220 ± 250 | 1130 ± 60 | **0.051 ± 0.005** | **0.19 ± 0.01** |
|  |  | 5.47 | 13.7 ± 1.8 | 0.016 ± 0.002 | 120 | 1060 ± 65 | 284 ± 17 | 0.21 ± 0.01 | 0.77 ± 0.04 |

### III. YH$_4$+YH$_6$ mixture in Uemura plot

From deduced ξ(0) values (Fig. 1 and Table 1), we calculated the Fermi temperature, $T_F$, for YH$_4$+YH$_6$ mixture by utilising standard approach of Bardeen-Cooper-Schrieffer theory [16] (details can be find elsewhere [17]):

$$T_F = \frac{\varepsilon_F}{k_B} = \frac{\pi^2}{8} \cdot m^*_{eff} \cdot \xi^2(0) \cdot \left(\frac{\alpha \cdot k_B \cdot T_c}{\hbar}\right)^2, \qquad (5)$$

where $\alpha = \frac{2 \cdot \Delta(0)}{k_B \cdot T_c}$, $\Delta(0)$ is the amplitude of the ground state energy gap, $\varepsilon_F$ is the Fermi energy, $\hbar = h/2\pi$ is reduced Planck constant, $k_B$ is the Boltzmann constant, $m^*_{eff}$ is the charge carrier effective mass ($m^*_{eff} = 2.73 \cdot m_e$ for YH$_6$ at $P$ = 200-350 GPa [11]).

For NRT superconductors α = 3.53-5.47, where the lower bound is reported for H$_3$S [17,18] and the upper bond reported for YH$_x$ [1]. Based on the fact that critical temperatures for YH$_4$+YH$_6$ mixture deduced by all four models are very close to each other, $T_c$ = 217.1 ± 0.4 K, in Table I we show only the $T_c/T_F$ ratios.

As the result, the mixture of YH$_4$+YH$_6$ ($P$ = 185 GPa) in all considered scenarios (Table 1) has $0.01 \leq T_c/T_F \leq 0.04$ and falls in unconventional superconductors band of the Uemura plot [19,20] in close proximity to other NRT counterparts [17],[21],[22] (Fig. 2).

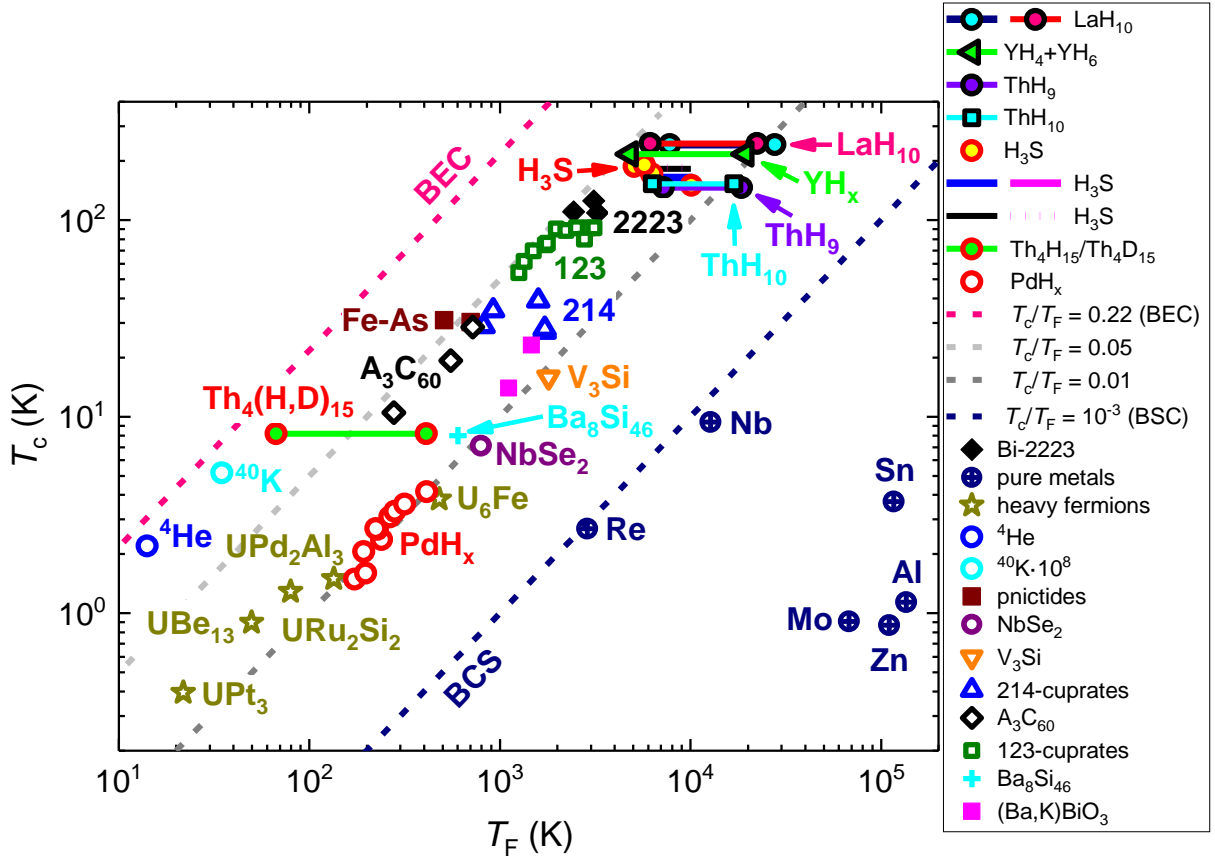

**Figure 2.** A plot of $T_c$ versus $T_F$ obtained for most representative superconducting families including PdH$_x$, Th$_4$H$_{15}$/Th$_4$D$_{15}$, ThH$_9$, ThH$_{10}$, H$_3$S, LaH$_{10}$, and the YH$_4$+YH$_6$ mixture (Sample 5 [2]). Other data was taken from [17],[19-22]. BEC and BCS lines are shown for clarity.

### IV. Thermodynamic fluctuations of order parameters of YH$_4$+YH$_6$ mixture

Both research groups [1,2] pointed out that experimental $T_c$ values for YH$_x$ superhydrides are about 40-50 K lower than values predicted by the first principles calculations. One possible reason for observed $T_c$ suppression can be originated by thermodynamic fluctuations of the order parameter [23,24]. There are two characteristic temperatures for thermodynamic fluctuations in superconductors, one is the phase order fluctuations temperature [23]:

$$T_{fluc,phase} = \frac{0.55 \cdot \phi_0^2}{\pi^{3/2} \cdot \mu_0 \cdot k_B} \cdot \frac{1}{\kappa^2 \cdot \xi(0)} \quad (6)$$

where $\kappa = \lambda(0)/\xi(0)$ is Ginzburg-Landau parameter, and $\lambda(0)$ is the ground state London penetration depth.

The second is the characteristic temperature for the amplitude order parameter fluctuations [24]:

$$T_{fluc,amp} = \frac{\phi_0^2}{12 \cdot \pi \cdot \mu_0 \cdot k_B} \cdot \frac{1}{\kappa^2 \cdot \xi(0)} \quad (7)$$

Calculated values for $T_{fluc,phase}$ and $T_{fluc,amp}$ are shown in Table I, where the value of $\kappa$ = 60-120 covers expected range for majority of high-temperature superconductors [17],[21],[25,26]. It can be seen, that in some scenarios $T_c/T_{fluc,amp}$ ~ 0.9, which means that observed in experiment the suppression in $T_c$ for the $YH_4+YH_6$ mixture of superhydrides can be explained by fundamental thermodynamic fluctuations.

## V. Results

In result, in this paper the mixture of NRT superconductors $YH_4+YH_6$ has been classified as unconventional superconductor which is nicely matched the location of other NRT superhydrides in the Uemura plot. It is also shown the thermodynamic fluctuations of the order parameter amplitude is dominating factor which limits superconducting transition temperature in superhydrides of yttrium.

## Acknowledgement

The author thanks financial support provided by the state assignment of Minobrnauki of Russia (theme "Pressure" No. AAAA-A18-118020190104-3) and by Act 211 Government of the Russian Federation, contract No. 02.A03.21.0006.